\documentclass[12pt]{article}

\usepackage{setspace,graphicx,epstopdf,amsmath,amsfonts,amssymb,amsthm,versionPO}
\usepackage{marginnote,datetime,enumitem,subfigure,rotating,fancyvrb}
\usepackage{hyperref,float}
\usepackage[longnamesfirst]{natbib}
\usdate
\usepackage[title]{appendix}

\excludeversion{notes}		% Include notes?
\includeversion{links}          % Turn hyperlinks on?

\iflinks{}{\hypersetup{draft=true}}

\ifnotes{%
\usepackage[margin=1in,paperwidth=10in,right=2.5in]{geometry}%
\usepackage[textwidth=1.4in,shadow,colorinlistoftodos]{todonotes}%
}{%
\usepackage[margin=1in]{geometry}%
\usepackage[disable]{todonotes}%
}

\makeatletter\let\chapter\@undefined\makeatother % Undefine \chapter for todonotes

\begin{document}

\setlist{noitemsep}  % Reduce space between list items (itemize, enumerate, etc.)
%\onehalfspacing      % Use 1.5 spacing
% Use endnotes instead of footnotes - redefine \footnote command

\title{Explanatory Power of Stocks Correlations}

\author{Ludovico Latmiral\footnote{ludovico.latmiral@hotmail.it
\newline
I am very grateful to the Multi-manager team of Kairos Investment Management Ltd for insightful discussions on the subject of this paper. In particular, I wish to thank Sabino Delfino for many discussions as well as for providing the data on the EURO STOXX market. 
The opinions and views expressed in this paper are uniquely those of the author, and do not necessarily represent those of Kairos Group.
The Author also declare no competing financial interest.}\\
Kairos Investment Management Ltd, London W1H 6AZ, United Kingdom}

\date{}              % No date for final submission

% Create title page with no page number

\renewcommand{\thefootnote}{\fnsymbol{ludovico.latmiral@hotmail.it}}

\singlespacing

\maketitle

\onehalfspacing

\vspace{-.2in}
\begin{abstract}
\noindent We analyze correlations among stock returns via a series of widely adopted parameters which we refer to as \textit{explanatory variables}. We subsequently exploit the results to propose a long only quantitative adaptive technique to construct a profitable portfolio of assets which exhibits minor drawdowns and higher recoveries than both an equally weighted and a \textit{standard} efficient frontier portfolio.
\end{abstract}

\medskip

% \noindent \textit{JEL classification}: XXX, YYY.

\medskip
% \noindent \textit{Keywords}: \LaTeX; papers with no content.

\thispagestyle{empty}

\clearpage

% \onehalfspacing
\setstretch{2}
\setcounter{footnote}{0}
\renewcommand{\thefootnote}{\arabic{footnote}}
\setcounter{page}{1}

\maketitle

\section{Introduction}

The links between financial ratios and assets returns have ever since been a fertile field of discussion in the finance industry \cite{welch2008}. Capital Asset Pricing Model (CAPM) is considered to be the first quantitative approach in this direction, assuming that regression coefficients of securities returns with respect to market returns (both in excess of the free-risk rate), \textit{i.e.} the $\beta_s$, are sufficient to describe the cross section of expected returns \cite{sharpe1964}.\\
Mathematically, betas can be obtained from the expression $\beta_{i,m}=\rho_{i,m}\ \sigma_i/\sigma_m$, where $\rho_{i,m}$ is the correlation between the $\mathrm{i}^{\mathrm{th}}$ asset and the market, $\sigma_i$ is the volatility of the $\mathrm{i}^{\mathrm{th}}$ asset and $\sigma_m$ is the market volatility. The assumption of a linear relation between expected returns and volatilities derives from the hypothesis of a symmetric (\textit{Gaussian}) distribution of assets returns and has induced the identification of risk with volatility.\\
Many studies have been subsequently devoted to identifying different kinds of \textit{risk premia}, investigating correlations between excess average returns and several possible factors of risk. Notorious examples are the negative and positive correlation of returns respectively with market capitalization ($\mathrm{CAP}$) \cite{banz1981} and with book-to-market ratio ($\mathrm{BtM}$) \cite{stattman1980}. E. Fama and K. French embedded these considerations in their seminal papers (see Ref.\cite{fama1992, fama2015}), where they proposed a three (and then a five) factor model based on $\beta$, $\mathrm{CAP}$ and $\mathrm{BtM}$ to explain the cross-sectional variation in average stock returns. Interestingly, the Authors also analyzed the relation of $\mathrm{BtM}$ and $\mathrm{CAP}$ with other widely used estimators. They provided evidence that $\mathrm{BtM}$ embeds both effects of book and market leverages, and together with $\mathrm{CAP}$ is able to fully explain the correlation of returns with the earnings-price ratio. Also, the small size of a company is shown to be generally linked with negative earnings and higher expected risks and returns.

In this paper we analyze the correlation of returns of stocks with a series of plausible \textit{explanatory variables}. We adopt parameters which are uniquely dependent on the historical price series such as volatility, skewness and beta with the market, and parameters which are linked to a bottom-up analysis of the company, \textit{e.g.} book and enterprise values, dividends, EBITDA.
The strength of our approach lies in finding insightful and durable relations within the market, which last also after the time frame needed to measure them. We have chosen to focus on correlations as they are strictly linked to the R square (which is simply the square of $\rho$), corresponding to the percentage variation of returns explained by the related \textit{explanatory variable}.\\
In the first part of our study, we regress the returns of each stock over the whole set of explanatory variables computed both over the same, contemporary, time window and over an antecedent time window. This analysis provides useful information on the correlations among different variables, offering possible clues on the time-dependent cause-effect relations between them.
Even though time delayed correlations generally reveal to be weaker, notably, we are able to identify an optimal subset of \textit{explanatory variables} which we prove to have a significant connection with future returns. The main result of our paper is precisely to propose a long-only\footnote{We focus on the construction of a long-only portfolio in order to reduce trading costs and avoid any complex (and debatable) modeling of shorts implementation costs.} asset allocation technique that exploits correlations across different time periods in various market environments over the last fifteen years. We will demonstrate the efficiency of our method by showing that it would have consistently outperformed equally weighted portfolios of stocks both in the American and in the European markets, also reporting a higher sharpe ratio (an equally weighted portfolio is \textit{per se} expected to outperform the corresponding capitalization based one because of the negative correlation of returns with the size of a company). Besides, the comparison with the equally weighted basket containing all the stocks at our disposal for selection provides generality to our methodology, removing the bias of working with a restricted subset of stocks.
Most importantly, our algorithmic approach also reveals a low momentum bias, being able to achieve higher performance in drawdowns and rebounds than a traditional efficient frontier allocation.

\section{Explanatory Fields}

We have based our analysis on Bloomberg historical daily data for a set of $100$ stocks picked from Standard and Poor's 500 index (SPX) and $360$ companies within Euro Stoxx index (SXXP). The selection process was almost random: we sorted companies ensuring to have sufficiently long historical track records as long as to obtain a trustful representative sample of SPX (and SXXP) in terms of the GICS sectors (see the Appendix for more details). Together with prices, downloaded data included the aforementioned market capitalization and book-to-market value, as well as the enterprise value (EV), the dividend yield (DivYield) and the EBITDA. We also considered a set of seven benchmark indexes: SPX, SXXP, Brent Crude Oil, VIX index (VIX), the generic US ten years government bond yield (US10Y) and the analogous Eurozone bond yield (EU10Y), and MSCI Value, Growth and Momentum indexes. We have computed correlations and $\beta_s$ for each stock under scrutiny with respect to these indexes. Hereafter we will mainly discuss results concerning the American market: full details on the comparative, identical investigation carried on Euro Stoxx Index are presented in the Appendix.\\
For the sake of a better readability, we now briefly summarize a list of the explanatory parameters we have considered\footnote{In the following we will always refer to returns as net of the risk-free rate, which we associate to the 2 years US treasury yield.}.

- The \textit{correlation} ($\rho_{s,b}$) between the stock under scrutiny and each of the benchmarks. This provides an estimate of the strength of the linear relationship between the two return time series. We also know that for linear regression $R^2_{s,b}:=\rho^2_{s,b}$, \textit{i.e.} the R squared, is the percentage of the price change of the stock that is explained by the benchmark.

- The \textit{beta} ($\beta_{s,b}=\rho_{s,b}\frac{\sigma_s}{\sigma_b}$) is equal to the correlation times the ratio between the standard deviations. It corresponds to the angular coefficient obtained with a least square linear regression of the asset returns versus the benchmark of reference.

- The \textit{past mean} is computed as the algebraic mean of an asset return series $x_{a,\tau}$ over a defined time interval in the past, \textit{i.e.} $X_a =\sum_{\tau} x_{a,\tau}$ with $\tau\in [-T,0]$. Depending on whether the market is following a momentum or a mean reversal trend, we expect it to be respectively positively or negatively correlated with the mean of future returns $X_a$ in the subsequent time frame $\tau \in [0,t]$.

- The \textit{volatility} ($\sigma_a=\mathrm{std}(x_{a,\tau})$) is commonly considered the best estimator for risk and a proxy for expected returns, accordingly with CAPM theory predictions. Our analysis will confirm this assumption, even though we will stress that, similarly to the past mean, it is a momentum biased predictor.

- The \textit{sharpe ratio} ($\mathrm{Sharpe}_a=X_a/\sigma_a$) characterizes how well the return of an asset compensates the investor for the risk taken. It is usually adopted as a benchmark to compare different assets.

Further to the above mentioned well known figures of merit, we devoted special attention to the skewness of the returns distribution and to correlations with correlations and betas with respect to meaningful benchmarks, such as Fama and French type of factors and cross-correlations.\\
Let us briefly discuss these quantities in the next subsections.

\subsection{The skewness}

In probability theory, the skewness of a distribution is defined as its third standardized moment
\begin{equation}
\zeta_a=\frac{\mathrm{E}[(x_{a,\tau}-X_a)^3]}{\sigma_a^3}\ ,
\end{equation}
and it measures the asymmetry of the returns about the mean. From a qualitative perspective, negative skew indicates that the tail on the left side is longer or fatter than the tail on the right side and viceversa. The quantity was proposed in Ref.\cite{lemperiere2016} as a better proxy for risk than the standard deviation. In that paper, Lemperiere et al. argued that the contribution of negative returns is mainly amenable to few severe drops, rather than to many small contributions.
The authors further suggest to adopt a revised version of the skewness (which we will indicate with $\zeta_a^*$) that better fits with their intent to discard small returns fluctuations around the mean and gauge the contribution of worst event scenarios.
They first normalize the return series with zero mean and unitary variance, then rank them by their absolute value and finally consider the area underneath the compounded curve.
Intuitively, what happens is that a large area is obtained when the average return is lowered by few very steep drawdowns, while there are many contributions slightly above the average which make the compounded graph grow fast since from the beginning.

\subsection{Cross Correlations}

A meaningful quantity that has been raising additional interest, especially since the large spread of Exchanged-Traded Funds (ETFs), is the cross correlation among various traded stocks \cite{pollet2010}. There is a vast amount of literature that has been dedicated to the study of asymmetric correlations, showing that stocks returns are more correlated in bear markets \cite{hong2006}, and generally the exploitation of spurious anomalous correlations is a field of great interest for many quantitative funds.
Here, we look at a synthetic index
\begin{equation}
\mathrm{P}(t)=\frac{1}{N^2}\sum_{pairs(i,j)}\rho_{i,j}(t)\ ,
\end{equation}
corresponding to the equally weighted average of all cross pair correlations among the set of stocks under scrutiny. Following the approach proposed by Fama and French \cite{fama1992}, we then exploit it as a benchmark to evaluate correlations and betas with the excess return time series for each stock.\\

\subsection{Correlation with $\rho$ and $\beta$}\label{corrofcorr}

An explanatory field that deserves particular attention is the correlation between the mean of an asset returns and the correlation (and/or the beta) of the return series itself with any relevant benchmark. Let us use the Greek letter $\eta$ to indicate alternatively $\rho$ or $\beta$. We have by definition
\begin{equation}\label{betabeta}
\rho_{\eta, X}=\frac{\sum_i(\eta_i-\bar\eta)(X_i-\bar{X})}{\sigma_X\sigma_\eta}\ ,
\end{equation}
where the index $i$ runs over all assets in the sample under analysis (\textit{e.g.} a portfolio), $X_i=T^{-1}\sum_\tau x_{i,\tau}$ ($\eta_i=T^{-1}\sum_\tau \eta_{i,\tau}$) is the average return over the time window under consideration $T$ of the return series $x_{i,\tau}$ ($\eta_{i,\tau}$) (here $\tau$ indicates the discrete time at which the return is calculated with respect to the asset value at time $\tau-1$). These correlations are most important as they are a measure of (\textit{i.e.} the square root of) the percentage change in the series $x_{i,\tau}$ that is explained by the series $\eta_{i,\tau}$. We trust this is an insightful measure as it is not directly biased towards the market performance in a given time period, but it rather reflects inner relations amongst price series. In particular, there is in principle no restriction to compute the time averages for $\eta_{i,\tau}$ and $x_{i,\tau}$ over the same time window. Besides, while it would be meaningless to regress daily time series with a time lag of several months (\textit{i.e.} to compute monthly lagged daily betas), we argue that we expect some factors, such as betas and volatility for example, to contain durable peculiar information on the related asset.\\
The understanding of this expression for $\eta=\beta$ is pretty straightforward. Indeed, we know that $\beta_i$ is defined as the best fit slope for the vector equation $x_{i,\tau}=\beta_iM_\tau +q$, where $M$ is the benchmark used for the regression and the expected value of the intercept $q$ is zero if the linear fit assumption is satisfied. $\beta_i$ can be computed via a least square minimization of the centred returns
\begin{equation}
\frac{d}{d\beta_i}\sum_\tau [(x_{i,\tau}-X_{i})-\beta_i(M_\tau-\bar{M})]^2 =0\ ,
\end{equation}
which leads to $X_{i}=\beta_i\bar{M}$. With this information at hand and substituting in Eq.\eqref{betabeta}, we expect $\rho_{\beta,X}$ to have the same sign of the average $\bar{M}$. This is a trivial though insightful consideration, meaning that the average return of an asset is positively (negatively) correlated with the $\beta$ with respect to its reference benchmark depending on the expected value of the benchmark itself.\\
On the other hand, there is no immediate interpretation of the correlation $\rho_{\rho,X}$, whose sign could generally be independent of the correlations $\rho$. For a qualitative understanding of the differences between $\rho_{\beta,X}$ and $\rho_{\rho,X}$, it is helpful to consider the contribution of the volatility, which is positively correlated with the returns and strongly correlated with $\beta$, though weakly and oppositely correlated with $\rho$.\\
In the following we will consider a set of widely adopted indexes as reference benchmarks, such as SPX, VIX, US10Y, MSCI US Momentum, Growth and Value. In addition, we will follow the approach suggested by Fama and French and adopt also a set of factors as markets of reference for our correlation analysis (\textit{e.g.} $\mathrm{EV/EBITDA}$, MtB, EV, DivYield). The procedure consists in sorting stocks in the market on the basis of these factors and then construct artificial indexes by subtracting the returns generated by stocks in the high rank of the index from those in the low part. We will show how these indexes play a significant role, especially when considering correlations with future returns.

\section{Correlations between explanatory fields and returns}\label{explanatoryfields}

\begin{figure}[!h]
\centering
\subfigure[]{
        \label{Corrbad}
        \includegraphics[width=1.0\textwidth]{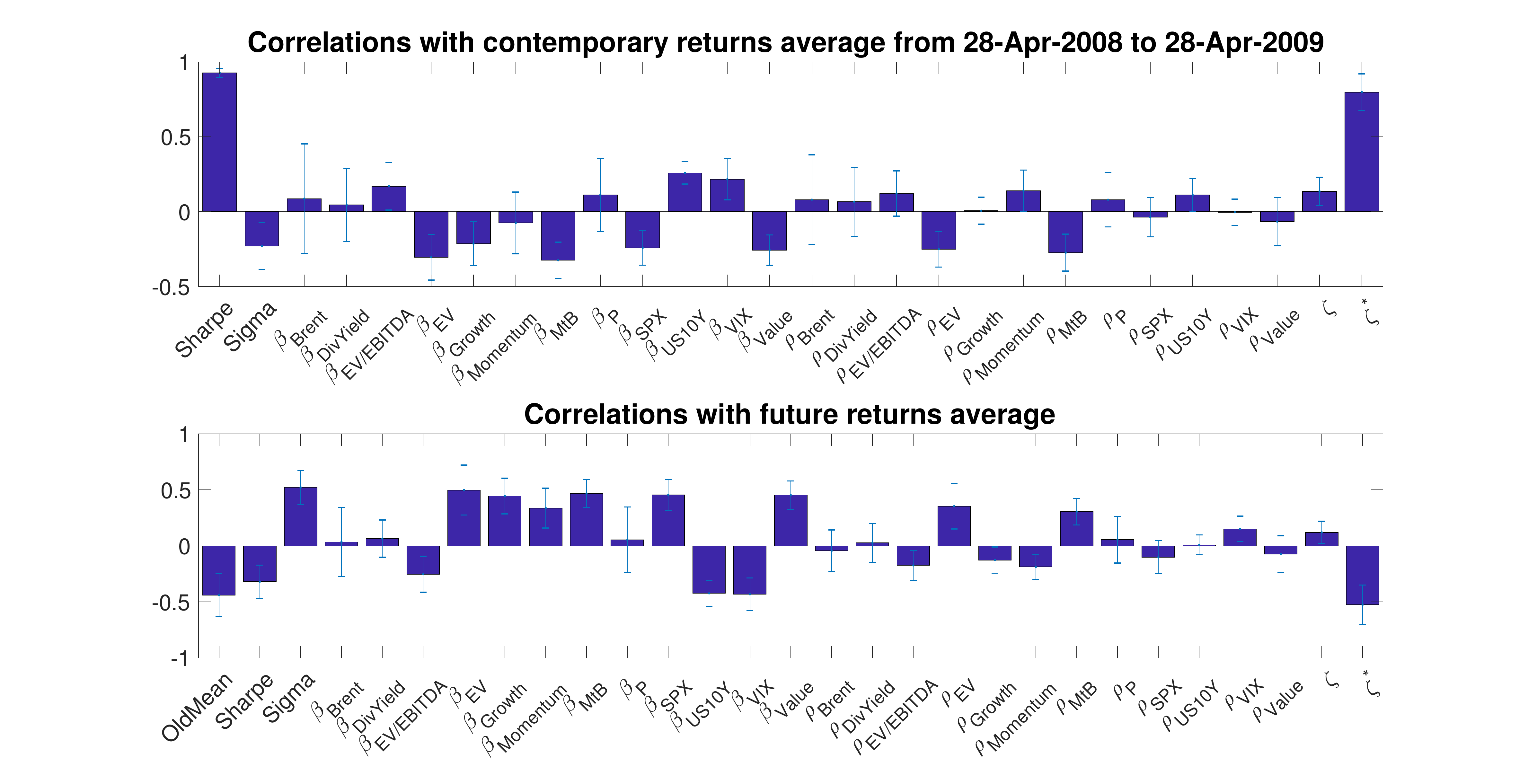} } 
\subfigure[]{
        \label{Corrgood}
        \includegraphics[width=1.0\textwidth]{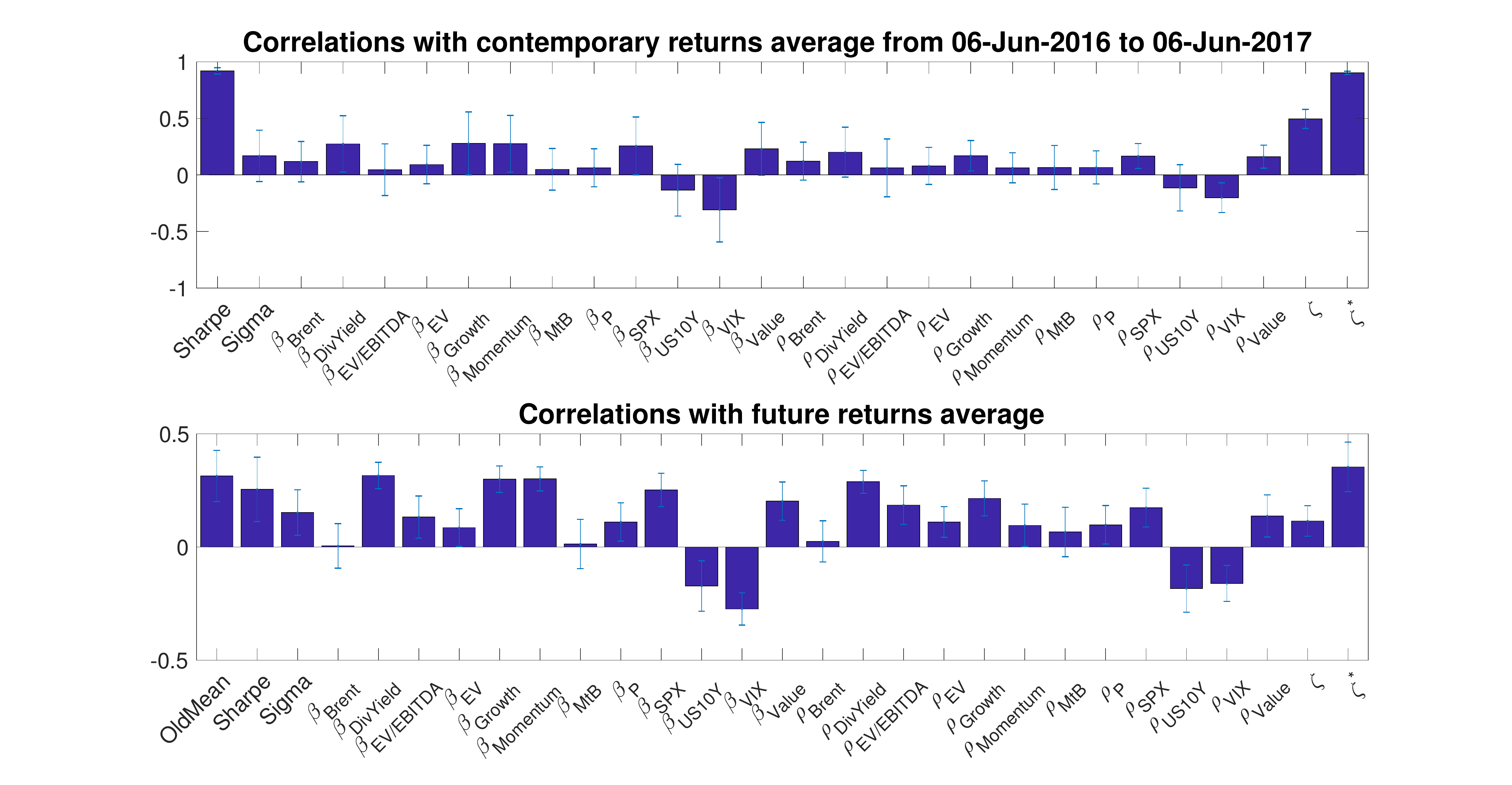} } 
\caption{Measured correlations of a set of explanatory fields with the average returns achieved in the same time window (upper diagrams) and in the following trading year (lower plots). Histograms (a) and (b) are referred, respectively, to a significantly volatile period (including the financial crisis in 2008) and a period of smoother growth and lower volatility. The analysis has been conducted over $100$ stocks picked from the $\mathrm{S\& P} 500$ index. Error bars indicate the $95\%$ confidence interval for each correlation.}\label{Correlations}
\end{figure}

Now that we have introduced the analytical tools and factors at the basis of our analysis, we present the first part of this paper, consisting in the study of correlations between a set of possible explanatory fields and return time series (see Fig.\ref{Correlations}).\\
We consider two time windows which were characterized by almost opposite market conditions. The former, from April 2008 to April 2009, embraced the last financial crisis and was marked by considerable volatility on the market and followed by a period of mean reversion and growth. The latter spans from June 2016 to June 2017 and refers to a period of almost stable growth with low volatility. In both cases we plot yearly correlations of a set of (possible) explanatory fields with contemporary returns and with returns obtained on a subsequent annual time window. While in Fig.\ref{Corrgood} the market conditions for the two years remained the same and all the correlations maintained their signs, the mean reversion that took place between June 2008 and June 2010 is reflected in a switch of sign in Fig.\ref{Corrbad} for all the most significant parameters. Actually, as discussed in Sec.\ref{corrofcorr}, the plot suggests that contemporary correlations with respect to volatility and to the most relevant betas change sign in drawdown market periods (with the exception of the correlation with the revised skewness $\zeta^*$). This is also confirmed by several identical analysis over different time windows embracing 2008 crisis).\\
We believe the comparison between the two time periods of both contemporary and time delayed correlations to be very insightful. Hereafter, we summarize the most relevant aspects in a point by point discussion.
\begin{itemize}
\item As expected by definition, the sharpe ratio is highly correlated with contemporary returns, irrespectively of the market environment. However, it is not a good estimator of future earnings, as it structurally lacks information to identify mean reversions.
\item The skewness of the returns distribution ($\zeta$), and in particular the adapted version proposed in Ref.\cite{lemperiere2016} ($\zeta^*$), reveals to be an excellent indicator of risk reward. When computed over the same time window, $\zeta^*$ grasps risk-return reward even more effectively than volatility. Most remarkably, it is positively and strongly correlated with contemporary returns also during the drawdown period April 2008 - April 2009, which indicates some bias with the average return. However, skewness seems to perform poorly in terms of future predictability.
\item While cross pairs correlations ($\rho_\mathrm{P}$) are generally supposed to rise in crisis environment, they have also risen in the last years of bull market, mostly because of the increasing role played by ETF instruments, pension funds, \textit{etc}. Nevertheless, they happen to be weakly related with stock returns in the periods under scrutiny and, at least in this format, they would not provide any significant predictive power.
\item The volatility $\mathrm{Sigma}$ ($\sigma$) is confirmed to be a highly relevant figure of merit, displaying high correlation with future returns series in a mean reversion environment. As a very important remark, however, it seems to be anti-correlated with contemporary returns in the drawdown period (see Fig.\ref{Corrbad}), indicating that volatile stocks perform poorly in bear markets, though they outperform in bull markets.
\item The correlation of returns with correlations with various explanatory factors $\rho_{\rho,X}$ does not account any significant relevance in neither the time periods, nor for contemporary or future returns.
\item Betas with respect to SPX, VIX (respectively the market to which the stocks under scrutiny belong to and the associated volatility index) and US treasuries show promising correlations with average stock returns.
\item MSCI Momentum, Growth and Value indexes also display significant correlations with future returns average and they maintain the same sign in both the market environments.
\item As expected from past literature, betas with respect to EV and MtB were negative in 2008-2009, while they are reverted when computed for lagged returns. We infer that this is because the largest companies that survived the crisis, and that were subsequently able to master financial leverage, were also those which outperformed in the recovery period. A point that should be further investigated is the decreasing in magnitude of the correlations with $\beta_{EV}$ and $\beta_{MtB}$ over years.
\item Eventually, the dividend yield and the ratio $\mathrm{EV/EBITDA}$ seem to have minor correlations with return series, the former acquiring larger significance in scenarios of stable growth (see Fig.\ref{Corrgood}).
\end{itemize}

\section{Time correlations with explanatory parameters}

The analysis on the correlations in the previous section reveals that observables measured in a certain time frame generally have little explanatory power on the dynamics in the subsequent time frame. On the other hand, it shows that many correlations depend on the market environment, which makes them trend biased.
Unfortunately, de-trending the return series does not guarantee a significant improvement in this direction and several attempts have failed to bring any further stability in the correlations, nor any significant insight on their dynamics.\\
As a matter of fact, it has historically been a hard task to foresee drawdowns in the market, and in those rare cases when this was possible, evidence was usually found in macro-economical or geo-political factors, rather than in historical time series \cite{kroencke2018}.
It is in this context that we trust an accurate and plain analysis of stock picking techniques, that could be applicable within different market environments, would be of high relevance. We believe that a reasonable convenience can derive from a change in perspective: instead of aiming at predicting future asset returns, we focus on forecasting the relationships amongst those returns. We do not attempt to investigate whether investing in a given market could be profitable in a certain macro economical and political environment. Instead, once an investment decision has been made (\textit{e.g} to take a long position in the US stock market), we try to address the problem of an efficient allocation of the resources.\\
In the following, we will present two possible approaches to deal with asset allocation and backtest their effectiveness from 2003 until today. The first method consists in the well known creation of a portfolio via the efficient frontier, while the second is a new proposal that takes into account the correlations analyzed in the previous section.\\
We will devote the next two subsections to the presentation of these methods, which are both based on a monthly reallocation of the stocks in the portfolio.

\begin{figure*}[t!]
\includegraphics[width=1.0\textwidth]{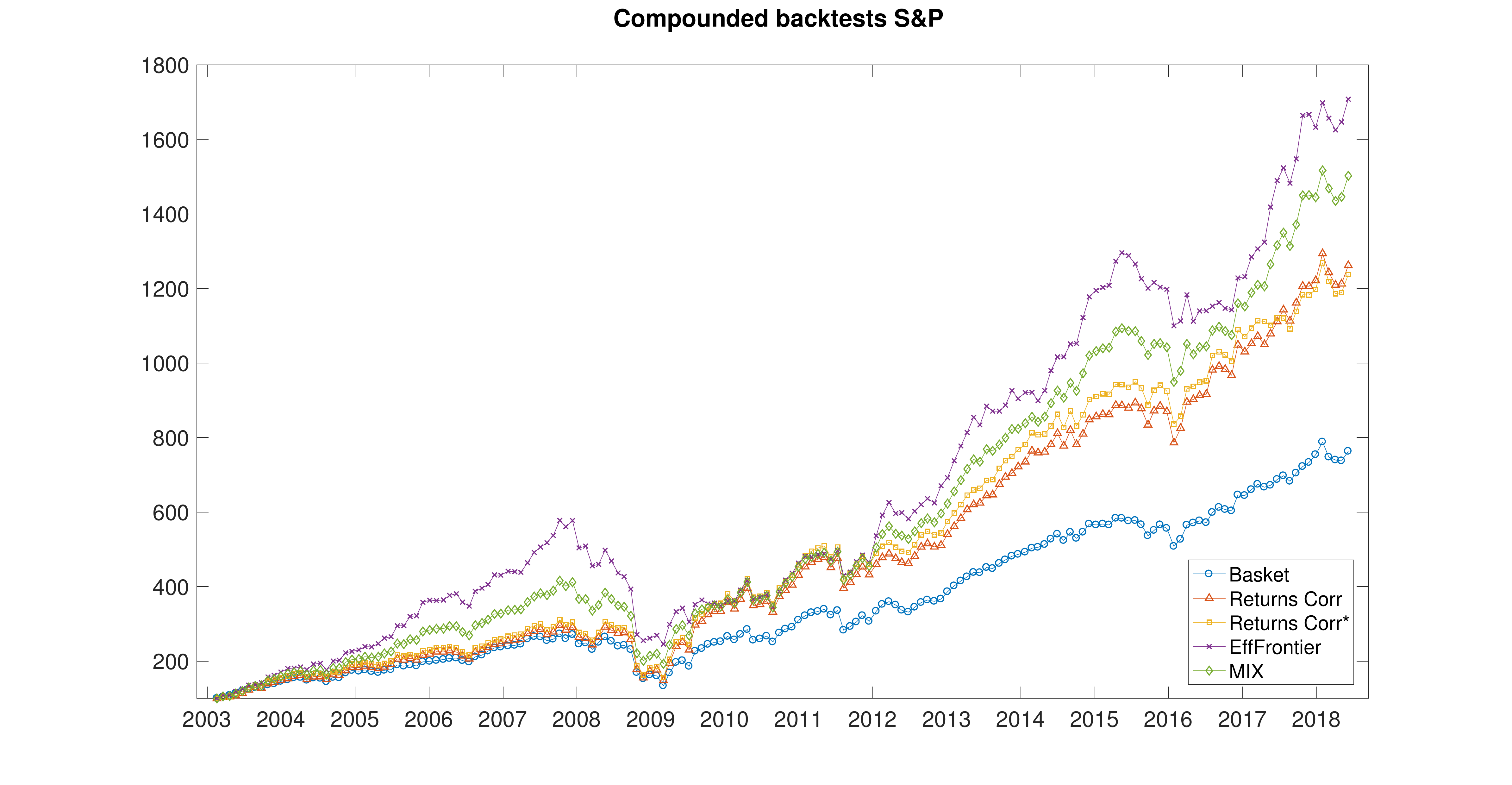}
\caption{Compounded monthly backtest of the discussed strategies from January 2003 to June 2018: monthly rebalanced equally weighted portfolio with 100 stocks (blue circles), allocation via efficient frontier ($EF$) (purple crosses), returns correlations ($RC$) (red triangles), mix strategy ($MIX$) (green rhombuses) and adaptive returns correlations (yellow squares).}\label{performances}
\end{figure*}

\subsection{Allocating via Efficient Frontier}

Given a set of assets, each with an historical average return and volatility, the efficient frontier is defined as the ensemble of all portfolios that minimize volatility for each expected return. In our backtested simulation, on each asset allocation date we considered the time series of the previous year and retrieved on this basis the efficient portfolio having the same volatility as the equally weighted portfolio containing all the stocks in our basket, but higher expected return. Resources were monthly allocated with the following simple rule: the portfolio was fully invested with equal weights only on those stocks whose proposed \textit{efficient} weights were larger than the average weight (\textit{i.e.} $w_i | w_i>1/N$, where $N=100$ is the total number of stocks and $w_i$ the proposed weight for the $\mathrm{i^{th}}$ asset). In the following, we will refer to this method as $EF$.

\subsection{Allocating via Returns Correlations}

Returns correlations are the main focus of this paper and we propose a technique to exploit them to construct effective portfolios (we will denote such protocol with the shortcut $RC$). The idea is to hinge every allocation on the regression of return time series of the previous year with respect to the explanatory fields measured over the second previous year, \textit{i.e.} with one year time lag with respect to the returns (overall, this method thus requires two years of backtest data). The multi-regression coefficients are then used, together with explanatory fields measured in the latest year, to infer returns for the incoming year. Resources are then fully and equally allocated in those assets whose predicted returns outperform the average predicted return of the whole set of $N$ stocks.\\
The effectiveness of this approach crucially depends on the ability to capture future returns through a set of explanatory fields. Thanks to the analysis presented in Sec.\ref{explanatoryfields}, we have been able to select the most significant and stable parameters. We have chosen ten factors: the revised skewness $\zeta^*$, the volatility and eight betas with respect to SPX, Momentum, Growth, $\mathrm{US10Y}$, VIX, DivYield, EV and MtB.\\

\section{Discussion}

We plot in Fig.\ref{performances} a backtest of the proposed strategies, together with a comparison with the equally weighted, monthly rebalanced portfolio (referred to as $EW$): both strategies under scrutiny significantly outperform $EW$. A mixed strategy ($MIX$), which allocates by averaging between $EF$ and $RC$, is shown to provide higher returns with limited volatility. A summary of the results with the most relevant figures of merit is presented in Table \ref{Tab1}. $RC$ outperforms $EW$ for $104$ months over $176$ (the period January 2003 - June 2018), which corresponds to around $\simeq 59\%$ of cases ($MIX$ reveals to be a winning strategy in $63\%$ of cases), and exhibits annualized sharpe ratio and return which are respectively $7.5\%$ and $26\%$ larger than for $EW$.\\
A figure of merit that deserves special attention from a statistical perspective is the confidence level with which our strategy identifies the outperforming stocks and can be distinguished from a random allocation of weights (we indicate it with \textit{Fidelity} in Table \ref{Tab1}). In other words, we demonstrate that our algorithm is truly able to infer some extra information on future returns distribution, which confers it effectiveness in the stock picking process. We compare each monthly asset allocation via $EF$ and $RC$ techniques with the true historical subsets of out- and under-performing stocks. If there was no predictive power, the logical $and$ comparison between forecasts and actual results would behave as a random binomial variable with probability $p$ equal to the probability that a stock randomly picked from a basket has greater returns than the average basket return. We expect $p\simeq 0.5$, with $p\rightarrow 0.5$ for large $N$. Satisfactorily, our approach is able to dismiss the random hypothesis, with $RC$ technique being slightly more precise than $EF$ at selecting the outperforming stocks with a confidence level of $87\%$.
This encouraging result is due to a difference in the fundamentals of the two procedures: while $RC$ allocation is expected to fulfill this precise scope, \textit{i.e.} to identify the subset of outperforming stocks, $EF$ approach does not necessarily select the assets with higher returns, but it accounts also for low volatility, generally resulting in a higher differentiation amongst stocks in the portfolio (see sharpe ratios and means in Table \ref{Tab1}).\\
There is another promising remark that Fig.\ref{performances} brings to our attention. As expected from its back-looking approach, $EF$ is strongly momentum biased and performs well in periods of stable growth, while it deeply underperforms during drawdowns. Instead, this bias is well handled by our proposal that uses returns correlations. On the one hand, $RC$ losses during the drawdowns in 2008, 2011 and 2015 are comparable with those of the $EW$ basket, significantly reducing the maximum drawdown registered by $EF$ (see Table \ref{Tab1}). On the other hand, $RC$ is extremely efficient at exploiting market rebounds after drawdowns: it largely outperforms $EF$ (and $EW$) both in the bear period 2008-2009 ($+10\%$) (and $+1\%$) and in the subsequent recovery 2009-2010 ($+66\%$) (and $+44\%$). A similar situation is replicated in 2015-2016 and 2016-2017.
This ability can be further enhanced capitalizing what we learned in Sec.\ref{explanatoryfields}. Indeed, we have observed that correlations of betas with contemporary returns are inverted in strongly bear markets, correlation with the volatility being the first indicator of such a reversion. Since market drawdowns usually last for a short period of time with respect to the whole economic cycle, when this condition is detected, we could adaptively revert the sign of the coefficients in the multi-regression analysis. This shrewdness allows to exploit the market rebound even better and boosts performance after a negative period ($+ 150\%$ return is achieved in 2009). The result is plotted in Fig.\ref{performances} and reported in Table \ref{Tab1} with the label ``\textit{Returns Corr*}''.

\begin{table}[t!]
\centering
\begin{tabular}{lcccccc}
\hline
 &
\multicolumn{1}{c}{Return} &
\multicolumn{1}{c}{Sharpe} &
\multicolumn{1}{c}{Max UP} &
\multicolumn{1}{c}{Max DD} &
\multicolumn{1}{c}{Months+} &
\multicolumn{1}{c}{Fidelity}\\
\hline
Eff Frontier   & 21.7\% & 1.08 & 81\% & -55\% & 102 & 0.83\\
Returns Corr   & 19.9\% & 0.91 & 147\% & -45\% & 106 & 0.87\\
Returns Corr*  & 19.8\% & 0.91 & 150\% & -45\% & 104 & 0.86\\ 
MIX            & 20.8\% & 1.03 & 99\% & -50\% & 108 & -\\
Equal Weights  & 15.8\% & 0.85 & 103\% & -46\% & - & -\\
\hline
\end{tabular}
\caption{Comparison of the investment strategies described in the text over the period January 2003 - June 2018. We have considered annualized returns, sharpe ratio and the number of months in which the strategy has outperformed the allocation with equal weights (\textit{Months+}). \textit{Max UP} (\textit{DD}) is the maximum annual gain (drawdown) and \textit{Fidelity} is the average fidelity (over the entire set of 176 months) with which the strategy can be distinguished from a random allocation.}
\label{Tab1}
\end{table}

\section{Conclusions}

Instead of looking at the cross section of stock returns, \textit{i.e.} trying to explain the dependency of historical time series on a set of contemporary benchmark time series, we have investigated correlations of (average) asset returns with respect to several explanatory fields, \textit{i.e.} quantities that are supposed to deliver information on assets performance across contiguous time intervals.\\
We have specifically looked at the correlations of returns with betas with respect to widely recognized financial indexes (SPX, VIX, Oil, US10Y, MSCI Momentum, Growth and Value) and factors (EV, DivYield, MtB, EBITDA). This approach was suggested by what we believed would be the interest of a portfolio manager who is always concerned about the future performance of his/her portfolio with respect to relevant benchmarks. We have chosen to concentrate on the time-lagged relationships among assets in a given basket with respect to several benchmarks factors.
We have furthered the study of different betas by looking at the explanatory power that past betas, volatility and skewness have on future average stock returns. We have thus defined a set of \textit{explanatory variables} which we have adaptively used to understand which assets were likely to outperform their basket of reference.\\
In order to provide evidence of the goodness of our approach we backtested our algorithm managing a long-only portfolio over a period of fifteen years (January 2003 - June 2018). The test was conducted on 100 stocks selected from $\mathrm{S \& P 500}$ index and on 360 stocks picked from Euro Stoxx index (numerical details on the latter are reported in the Appendix): very similar, satisfactory results were obtained in both cases.
On the one hand, our method managed to achieve higher sharpe ratio and higher overall annual performance than that of an equally weighted portfolio  ($27\%$ for SPX and $24\%$ for SXXP). On the other hand, it demonstrated a very low momentum bias both in the American and the European market, being able to deal well with 2008-2009 market crisis, reducing the drawdown experienced by $EF$ by $+10\%$, while increasing the rebound by $+66\%$.\\
We trust our approach could provide a concrete and resourceful benchmark for portfolio managers who deal with factor investing, while offering a platform for further studies on the explanatory power of many synthetic indexes, integrating very well with smart beta portfolio managing techniques.

\clearpage

\bibliographystyle{apalike}
\bibliography{Complete}

\clearpage

\begin{appendices}

\section{Basket Composition}

\begin{figure}[h!]
\centering
\includegraphics[width=0.65\textwidth]{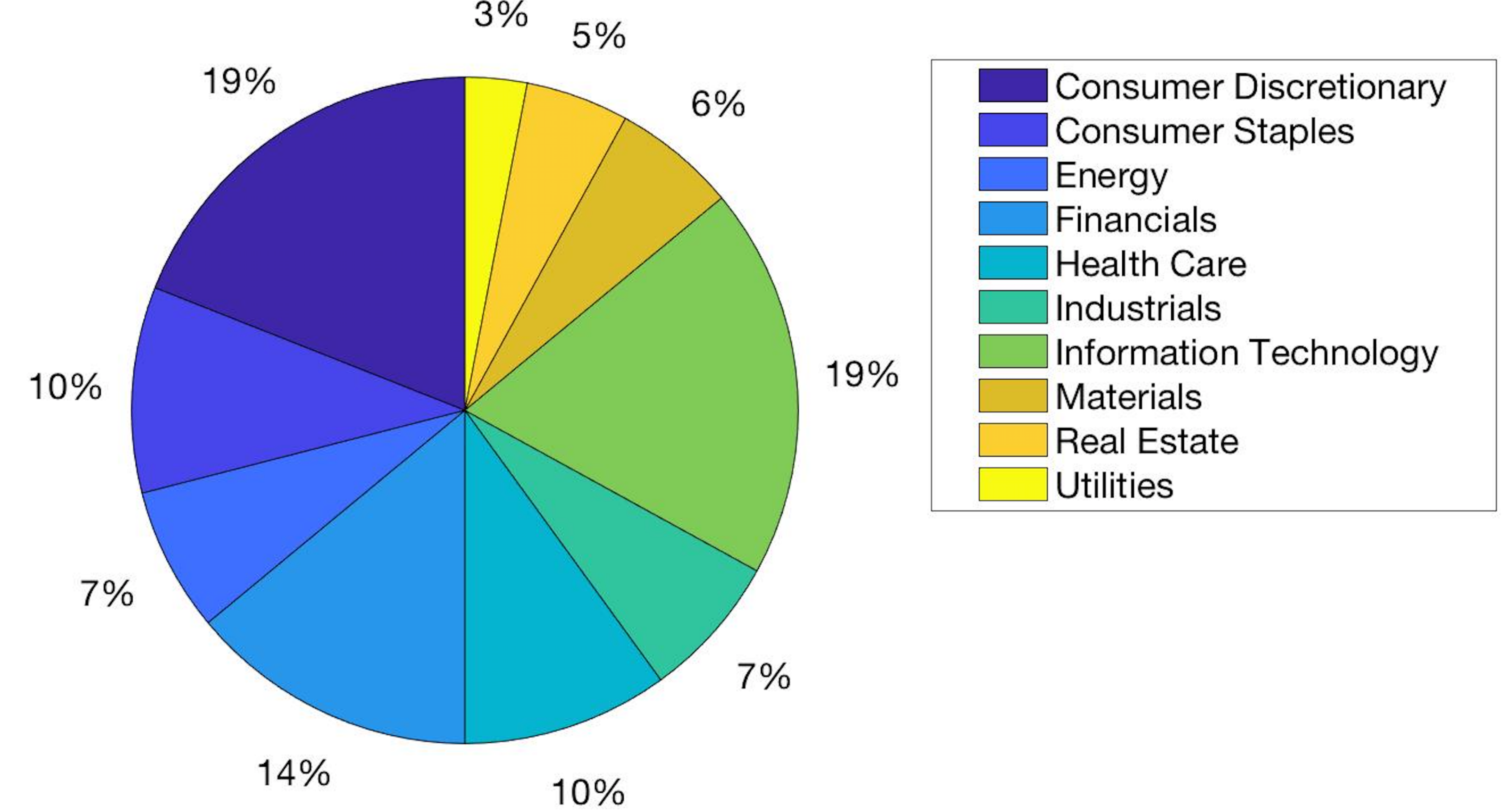}
\caption{Equally weighted basket composition in terms of GICS factors.}\label{piechart}
\end{figure}

We report in this section further details on the basket of 100 stocks that was used for the analysis (we would be pleased to provide the full list on demand, we simply refrain here for the sake of compactness).\\
The pie chart in Fig.\ref{piechart} shows the composition of the basket (equally weighted) in terms of GICS sectors. Even though there is no fundamental reason for which our selection of stocks should replicate $\mathrm{S \& P 500}$ index, a strong correlation with the index would witness the diversification of our basket and the generality of our results. We thus computed the return series of a basket containing our stocks and arranged with equal weights on 1st of January 2002 (with no subsequent rebalancements) and satisfactorily measured a correlation of $0.967$ with $\mathrm{S \& P 500}$ index over the time range until June 2018. 

\section{The European case}

Hereafter we present the results of the application of the analysis discussed in the main text on a basket of 360 companies listed on the Euro Stoxx index (SXXP).\\ Substantially, we observe similar correlations of stocks returns with our set of explanatory fields (we substituted SPX and US treasuries with SXXP and European generic 10 years bonds), and most satisfactorily we end up with a much more efficient and profitable portfolio management with higher annualized return ($+24\%$) and sharpe ratio ($+9\%$), with respect to an equally weighted portfolio with all the 360 stocks. As for the SPX, also in this case it is worth mentioning that the equally weighted portfolio already largely outperforms the capitalization weighted index. This is a common characteristic which can be explained by the negative correlation of returns with the size of a company (\textit{i.e.} its market capitalization).

\begin{figure}[t!]
\includegraphics[width=1.0\textwidth]{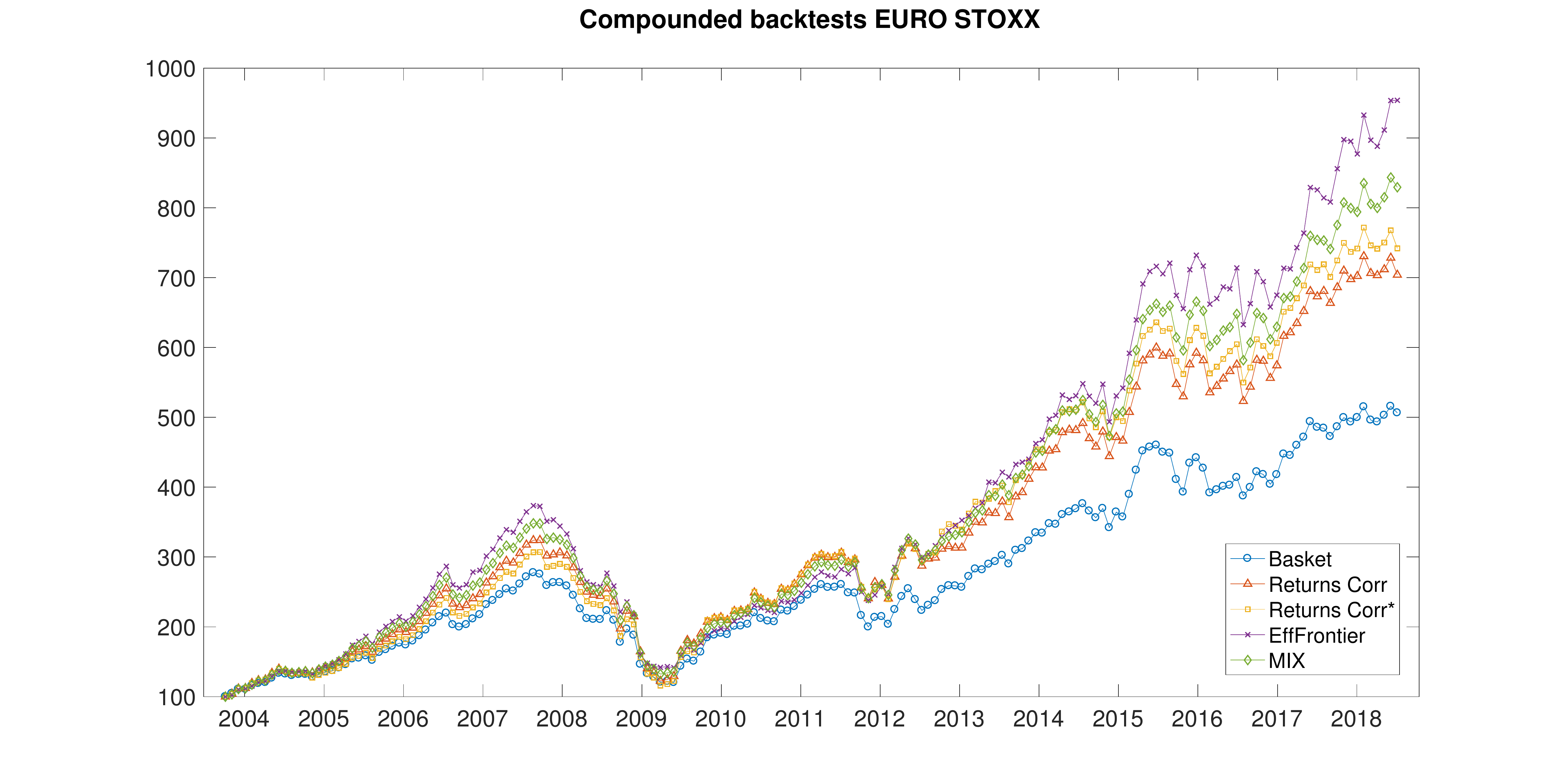}
\caption{Compounded monthly backtest of the discussed strategies from January 2003 to June 2018: monthly rebalanced equally weighted portfolio with 360 stocks from SXXP (blue circles), allocation via efficient frontier ($EF$) (purple crosses), returns correlations ($RC$) (red triangles), mix strategy ($MIX$) (green rhombuses) and adaptive returns correlations (yellow squares).}\label{performances}
\end{figure}

\begin{table}[b!]
\centering
\begin{tabular}{lcccccc}
\hline
 &
\multicolumn{1}{c}{Return} &
\multicolumn{1}{c}{Sharpe} &
\multicolumn{1}{c}{Max UP} &
\multicolumn{1}{c}{Max DD} &
\multicolumn{1}{c}{Months+} &
\multicolumn{1}{c}{Fidelity}\\
\hline
Eff Frontier   & 17.2\% & 0.96 & 63\% & -56\% & 118 & 0.70\\
Returns Corr   & 15.5\% & 0.79 & 93\% & -53\% & 105 & 0.80\\
Returns Corr*  & 15.8\% & 0.80 & 104\% & -53\% & 103 & 0.79\\ 
MIX            & 16.3\% & 0.90 & 78\% & -54\% & 109 & -\\
Equal Weights  & 12.7\% & 0.74 & 83\% & -49\% & - & -\\
\hline
\end{tabular}
\caption{Comparison of the investment strategies described in the text over the period January 2003 - June 2018 on 360 companies picked from SXXP. We have considered annualized returns, sharpe ratio and the number of months in which the strategy has outperformed the allocation with equal weights (\textit{Months+}). \textit{Max UP} (\textit{DD}) is the maximum annual gain (drawdown) and \textit{Fidelity} is the average fidelity (over the entire set of 176 months) with which the strategy can be distinguished from a random allocation.}
\label{Tab1}
\end{table}

\begin{figure}[ht!]
\centering
\subfigure[]{
        \label{Corrbadeu}
        \includegraphics[width=1.0\textwidth]{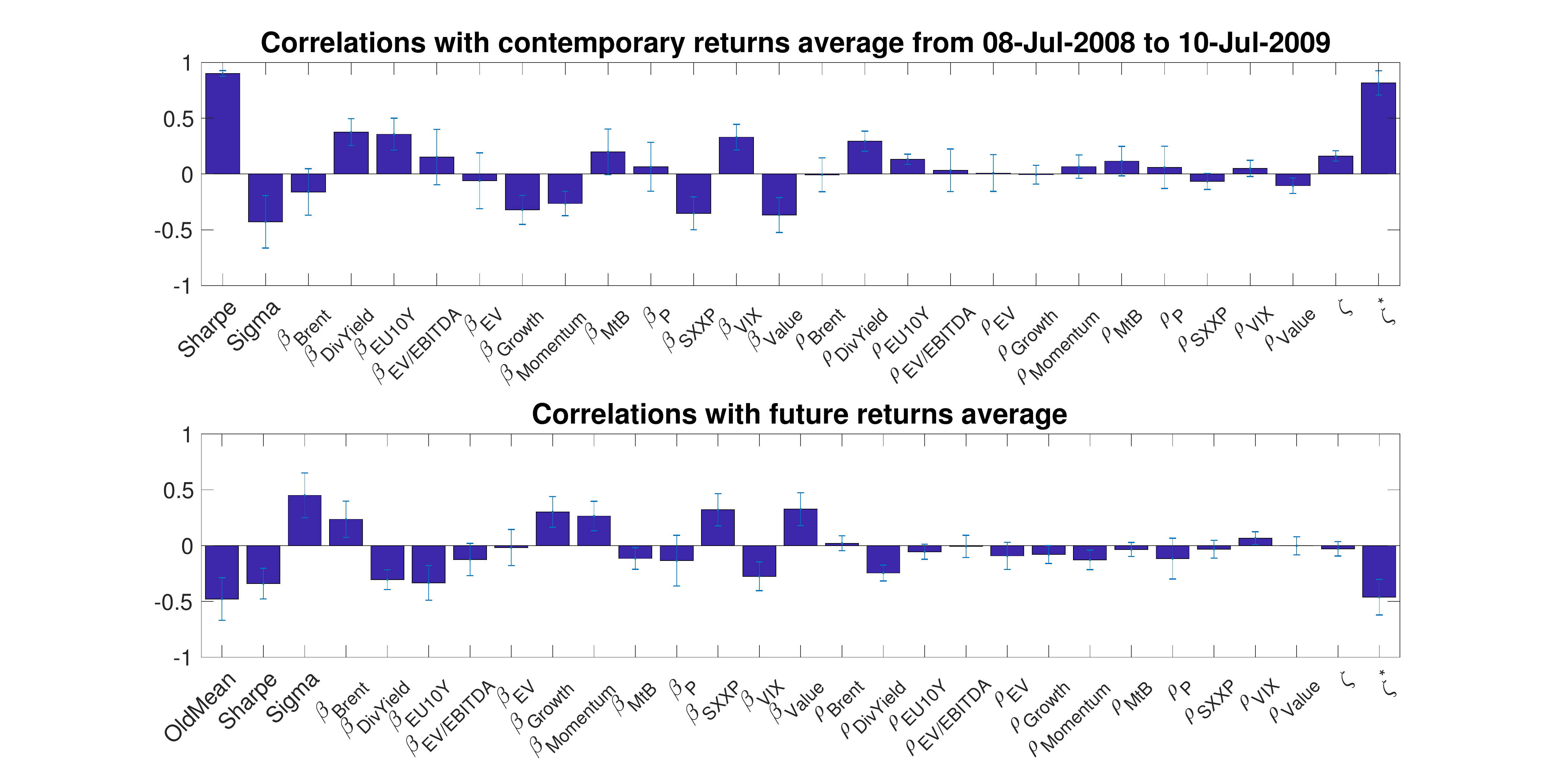} } 
\subfigure[]{
        \label{Corrgoodeu}
        \includegraphics[width=1.0\textwidth]{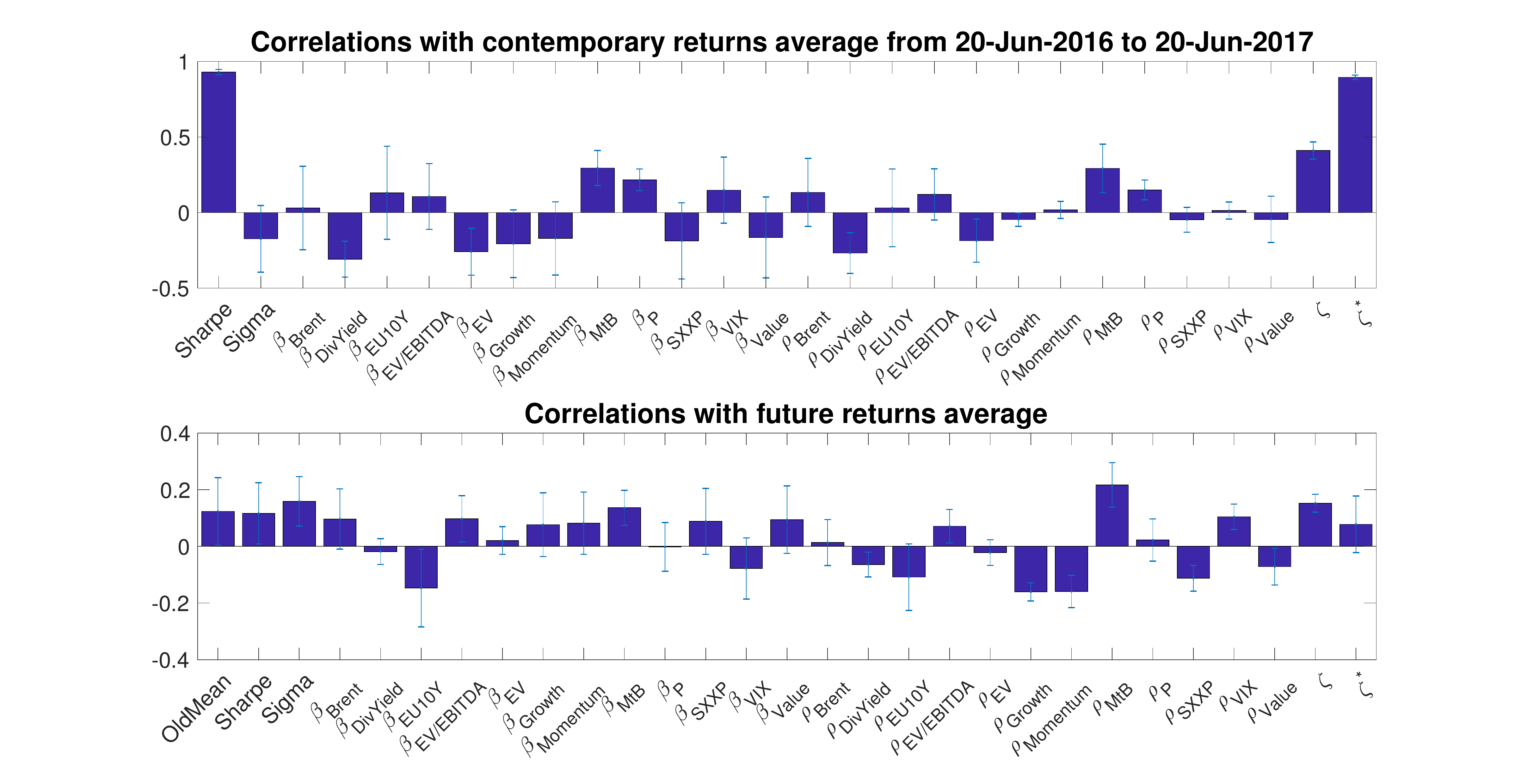} } 
\caption{Measured correlations of a set of explanatory fields with the average returns achieved in the same time window (upper diagrams) and in the following trading year (lower plots). Histograms (a) and (b) are referred, respectively, to a significantly volatile period (including the financial crisis in 2008) and a period of smoother growth and lower volatility. The analysis has been conducted over $360$ stocks picked from the Euro Stoxx index. Error bars indicate the $95\%$ confidence interval for each correlation.}\label{Correlations}
\end{figure}

\end{appendices}

%It may be useful to cite in this context Refs. \cite{black1986, bouchaud2017} where the Authors derive, under the hypothesis of efficient market, price oscillations of a factor of two around the \textit{correct} value of a tradable asset.

\end{document}